\documentclass[aps,prl,twocolumn,groupedaddress,showpacs]{revtex4}
\usepackage{graphicx}

\begin{document}

\title{Ultraviolet photonic crystal laser}
\author{X. Wu, A. Yamilov, X. Liu, S. Li, V. P. Dravid, R.  P. H. Chang and H. Cao}
\email{h-cao@northwestern.edu}
\affiliation{Material Research Center, Northwestern University, Evanston, IL, 60208}

\begin{abstract}
We fabricated two dimensional photonic crystal structures in zinc oxide films with focused ion beam etching. Lasing is realized in the near ultraviolet frequency at room temperature under optical pumping. From the measurement of lasing frequency and spatial profile of the lasing modes, as well as the photonic band structure calculation, we conclude that lasing occurs in the strongly localized defect modes near the edges of photonic band gap. These defect modes originate from the structure disorder unintentionally introduced during the fabrication process.
\end{abstract}

\maketitle


Photonic crystal slabs (PhCS), i.e., two-dimensional (2D) photonic crystals with finite vertical dimension, have attracted much attention because of their potential applications to various optoelectronic devices and circuits \cite{soukoulis}. Defect cavities in these photonic devices can have high quality factor and small modal volume\cite{Johnson:2001b,Vuckovic:2001,Srinivasan:2002,Akahane:2003}. The in-plane confinement is achieved via Bragg scattering, while the index guiding prevents light leakage in the perpendicular direction. Low-threshold lasing has been realized in PhCS made of III-V semiconductors.\cite{Painter:1999,Imada:1999,Benisty:1999,Hwang:2000} They operate in the near infrared frequencies. Ultraviolet (UV) photonic crystal laser, however, has not been realized yet. This is because shorter wavelength requires smaller feature size, which is technologically challenging for commonly used wide band gap materials (e.g., GaN, ZnO). On the other hand, the demand for blue and UV compact laser sources has prompted enormous research effort into wide band gap semiconductors. Compared with other wide band gap materials, ZnO has the advantage of large exciton binding energy ($\sim 60$ meV), that allows efficient excitonic emission even at room temperature.

In this letter, we realized, for the first time, ZnO photonic crystal lasers operating in the near-UV frequency at room temperature. We developed a procedure to fabricate 2D periodic structures in ZnO films with the focused ion beam (FIB) etching. Post thermal annealing is employed to remove structural damage induced by FIB etching. Lasing is achieved in the strongly localized defect modes near the edges of photonic band gap by optical pumping.  To explain the experimental results, we calculated the band structure of our samples using the 3D plane wave expansion method \cite{Johnson:2001}.

\begin{figure}
\vskip 0cm
\centerline{\rotatebox{0}{\scalebox{1.0}{\includegraphics{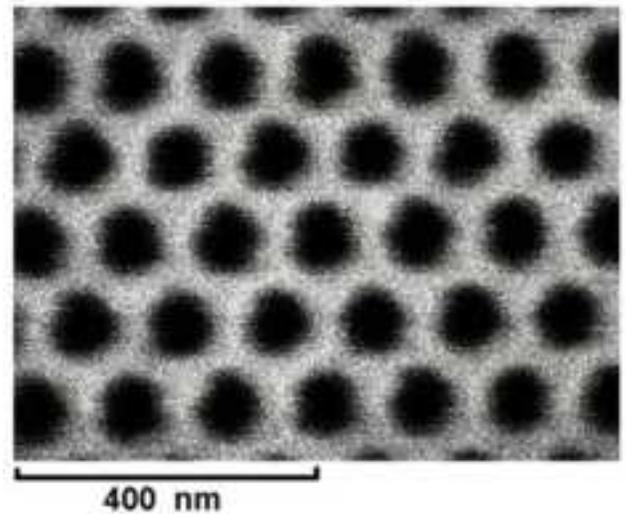}}}}
\caption{\label{sem} Top-view SEM of a triangular lattice ZnO photonic crystal slab. The lattice constant $a$ = 130 nm and air cylinder radius $r$ = 33 nm. }
\end{figure}

The 2D triangular lattice photonic crystal structure is fabricated in ZnO. First, 200 nm thick ZnO films are grown on c-plane sapphire substrates by plasma enhanced MOCVD at $750^{o}$ C. The selected area electron diffraction pattern of the film reveals that single crystalline ZnO is grown along the c-axis. Next, arrays of cylindrical columns are removed in the films by focused $Ga^{3+}$ ion beam etching at 30 KeV. FIB has been widely used for maskless and resistless nano-scale patterning, it allows us to precisely control the position, size and density of air cylinders in the ZnO films.  However, due to structural damage caused by FIB, the ZnO emission is quenched. To remove the structural damage, we annealed the patterned films in $O_{2}$ at $600^{o}C$ for one hour. To overlap the ZnO gain spectrum with the photonic band gap, the lattice constant $a$ and the radius of the air cylinders $r$ are varied over a wide range. The resolution of our FIB system limits the smallest step of length variation to 15 nm. The size of each pattern is about 8$\times$8 $\mu m$ with roughly 4000 air cylinders. A top-view scanning electron micrograph (SEM) of a part of one pattern is shown in Fig. \ref{sem} ($a$=130 nm and $r/a$=0.25). Side view SEM (not shown) reveals the air cylinders are etched through the ZnO film with nearly vertical walls. 

\begin{figure}
\centerline{\rotatebox{0}{\scalebox{0.33}{\includegraphics{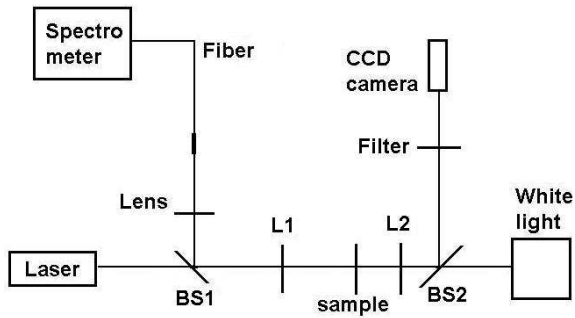}}}}
\caption{\label{setup} Schematic of the setup of our optical measurement. BS1 and BS2 are beam splitters, L1 is 10x objective lens, L2 is 20x objective lens.}
\end{figure}

The samples are optically pumped by the third harmonics of a mode-locked Nd:YAG laser (355 nm, 10 Hz, 20 ps) at room temperature. A schematic sketch of the experimental setup is shown in Fig. \ref{setup}. A 10$\times$ microscope objective lens (N.A.=0.25) is used to focus the pump beam to a 4 $\mu$m spot on one pattern, and also collect the emission from the pattern. Then the emitted light is focused by another lens into a UV fiber, which is connected to a spectrometer with 0.13 nm spectral resolution. Since the sapphire substrate is double-side polished and transparent in both visible and UV frequencies, a 20$\times$ microscope objective lens (N.A.=0.40) is placed at the back side of the sample for simultaneous measurement of the spatial distribution of emission intensity. The pump light is blocked by a bandpass filter, while the image of lasing mode profile is projected by the objective lens onto a UV sensitive CCD camera. The sample is also illuminated by a white light source so that we can identify the position of the lasing modes in the photonic lattice. 

\begin{figure}
\centerline{\rotatebox{0}{\scalebox{0.5}{\includegraphics{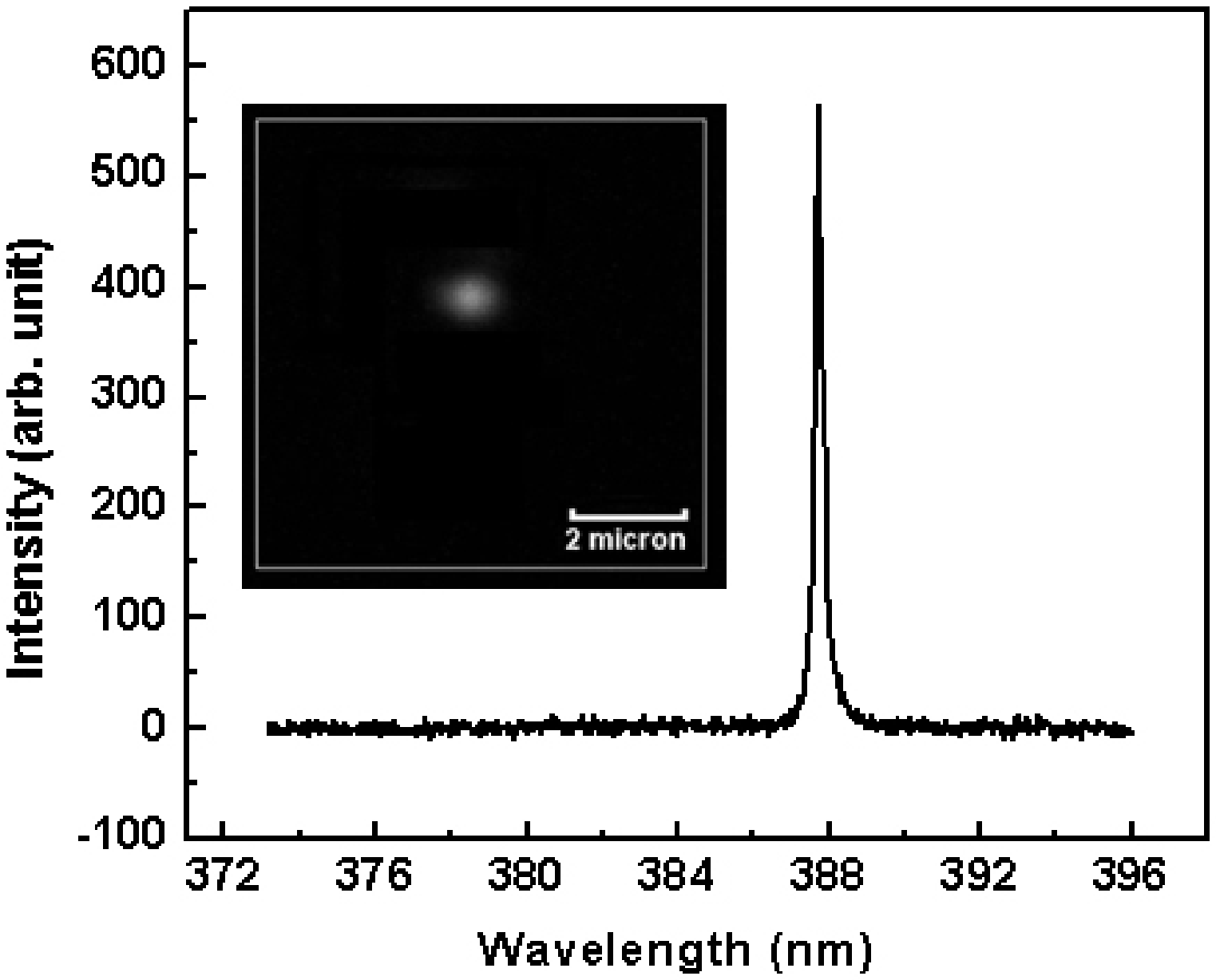}}}}
\centerline{\rotatebox{0}{\scalebox{0.85}{\includegraphics{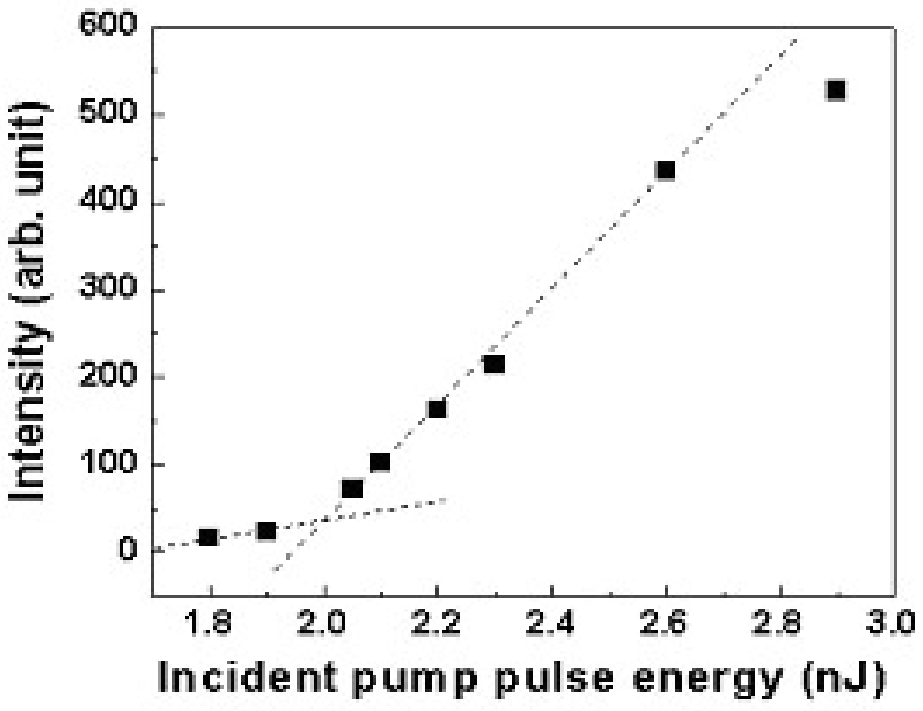}}}}
\caption{\label{specimg} (a): Lasing spectrum of a ZnO photonic crystal with $a$ = 115 nm shows a single defect mode. The incident pump pulse energy is 2.3 nJ. The inset is the near field image of this lasing mode. (b): Emission intensity of the defect mode versus the incident pump pulse energy.}
\end{figure}

Among all fabricated patterns with lattice constant $a$ varying from 100 to 160 nm, lasing is realized only in the structures of $a$ = 115 nm and 130 nm. Fig. \ref{specimg}(a) shows the spectrum of emission from a pattern of $a$ =115 nm and $r/a$ = 0.25. It has a single sharp peak at 387.7 nm. Fig. \ref{specimg}(b) plots the emission intensity integrated over this peak as a function of the incident pump pulse energy. The threshold behavior is clearly seen. Above the threshold, the spectral width of this lasing peak is only 0.24 nm.  These data indicate that lasing oscillation occurs in this structure. The near-field image of this lasing mode is obtained simultaneously and shown in the inset of Fig. \ref{specimg}(a). The white square marks the boundary of the triangular lattice. The lasing mode is spatially localized in a small region of $\sim$1.0 $\mu m^2$ inside the lattice.  As we move the pump spot across the lattice, the lasing modes change in both frequency and spatial pattern. This behavior suggests that the lasing modes are spatially localized defect states. They are formed by the short range structural disorder\cite{disphc}, not ``stacking faults''\cite{VardenyPhotonicLasing}, which perturb the long range periodicity. In the SEM (Fig. \ref{sem}), non-uniformity of the size and shape of air cylinders is quite evident. Lasing is also observed in the lattices of $a$=130 nm and $r/a$ = 0.25. However, the lasing modes have longer wavelength, and their lasing threshold is higher. 
  
\begin{figure}
\centerline{\rotatebox{-90}{\scalebox{0.33}{\includegraphics{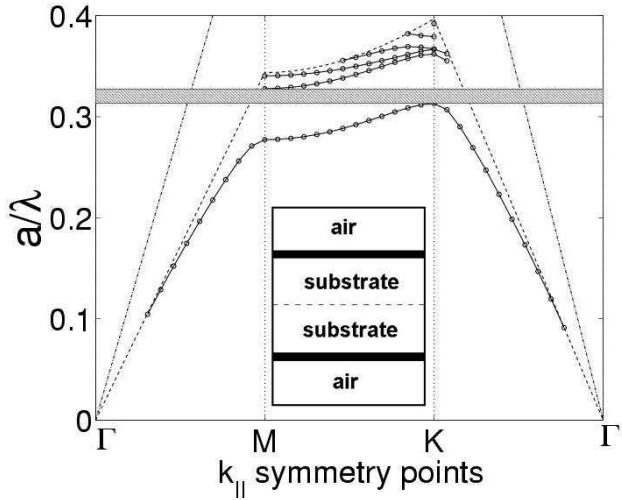}}}}
\caption{\label{cal} Calculated band structure of ZnO photonic crystal slab with lattice constant $a$ = 130 nm, air cylinder radius $r$ = 0.25$a$, and slab thickness $t$ = 200 nm. The refractive indices for ZnO and sapphire are 2.34 and 1.68. Dotted line is light cone between ZnO slab and air, dashed line is light cone between ZnO slab and sapphire substrate. Inset shows the super unit cell used in our calculation, the dark strips are ZnO photonic crystal slabs.}
\end{figure}

To understand our experimental results, we calculated the photonic band structures using the three-dimensional (3D) plane wave expansion method\cite{Johnson:2001}. Our ZnO PhCS is vertically asymmetric: with air on top and sapphire substrate at bottom. To apply the plane wave expansion method for photonic band calculation, we considered a super unit cell that consists of air, ZnO photonic crystal slab, sapphire substrate, ZnO photonic crystal slab and air (see the inset of Fig. \ref{cal}). The guided modes confined to the PhCS should lie inside the ZnO/substrate light cone and the air/ZnO light cone. If the substrate is thick enough, the guided modes in the two ZnO layers are decoupled, the symmetric and antisymmetric (with respect to the cell center plane -- dashed line in the inset of Fig. \ref{cal}) modes should be degenerate in eigenfrequency. This artificial degeneracy is used as the check of consistency in our band structure calculations. In a symmetric PhCS (i.e. the substrate is replaced by air), the modes can be grouped into two classes: modes that have z-component of the magnetic field or electric field symmetric with respect to the center plane of the photonic layer. The complete stop bands for the guided modes of each class can exist independently\cite{Johnson:1999}. Low lying modes of the first/second class are mainly transverse electric/magnetic (TE/TM) polarized.  The presence of the substrate removes the symmetry\cite{Chow:2000}. For our ZnO samples, we calculated the field components of the first five bands with and without the sapphire substrate and found close resemblance in the spatial profiles of the corresponding modes in the two cases. This is due to strong vertical confinement of the guided modes. Therefore, the modes in the ZnO PhCS with the sapphire substrate can still be classified as predominantly either TE-like or  TM-like. It is known that the polarization of ZnO exciton emission is mostly perpendicular to the c-axis\cite{Park:1966,Jung:2002}. Since the c-axis of our ZnO film is normal to the film/substrate interface, the emitted photons are coupled mainly into TE-like modes, as we confirmed experimentally from the measurement of polarization of photoluminescence from the side of an  unpatterned ZnO film. Therefore, we can disregard the TM-like bands in the band structure calculation. For an appropriate choice of lattice parameters, a complete photonic band gap exists between the first two TE-like bands\cite{later}. Fig. \ref{cal} shows an example of the calculated photonic band structures for $a$ =130 nm and $r/a$ = 0.25. For this system there exist a photonic band gap from 396 nm to 415 nm. For PhCS with $a$ =115 nm and $r/a$ = 0.25, a narrower gap exists between 363 nm and 372 nm. At room temperature, the ZnO gain spectrum is centered around 385 nm with a width of $\sim$ 12 nm. Thus the calculated photonic band gaps for the above structures do not exactly overlap with the gain spectrum. However, the unavoidable imperfections in the structure created during the fabrication will broaden the band gap and make it shallower \cite{band}. The spatially localized defect modes are formed near the band edges. We believe the lasing modes in our $a$ = 115nm structure correspond to the defect modes near the dielectric band edge (on the lower frequency side of the band gap), while the lasing modes in the $a$ = 130nm structure are the defect modes near the air band edge (on the higher frequency side of the band gap). The lasing threshold of the former structure is lower than that of the latter, because the defect modes at the dielectric band edge are concentrated inside ZnO and experience more gain. For all other systems, either there is no full band gap for TE-like modes or the ZnO gain spectrum is far away from the gap. Hence the lasing threshold is too high to be reached experimentally. 

In summary, we demonstrated optically pumped ZnO photonic crystal lasers. They operate in the near UV frequency at room temperature. The lasing modes are spatially localized defect states near the edge of photonic band gap, that are formed by short range structural disorder unintentionally introduced during the fabrication process. Further reduction of lasing threshold can be obtained by fine tuning of the structure parameters. 

This work was supported by the National Science Foundation under the grant no. ECS-0244457.

\end{document}